\documentclass[aps,prd,twocolumn,superscriptaddress,nofootinbib,showpacs,preprintnumbers]{revtex4-1}
\usepackage{amsmath}
\usepackage{slashed}
\usepackage{amssymb}
\usepackage{graphicx}
\usepackage{subfigure}
\usepackage{color}
\usepackage{hyperref}
\usepackage[normalem]{ulem}
\usepackage[T1]{fontenc}
\newcommand{\eq}[1]{Eq.~\eqref{eq:#1}}

\newcommand{\lb}{\Big{\lbrack}}
\newcommand{\rb}{\Big{\rbrack}}
\newcommand{\lp}{\Big{(}}
\newcommand{\rp}{\Big{)}}

\newcommand{\nn}{\nonumber}

\newcommand{\corr}[1]{{#1}_{\text{corr}}}

\newcommand{\NLLp}{NLL$'$}

\newcommand{\es}{e_\text{S}}
\newcommand{\eb}{e_\text{B}}
\definecolor{orange}{rgb}{1,0.5,0}
\definecolor{bluep}{rgb}{0.467,0.62,0.95}
\definecolor{bluep}{rgb}{0,0,0.9}

\begin{document}

\title{Subtracted Cumulants: Mitigating Large Background in Jet Substructure}

\author{Yang-Ting Chien}
\email[]{ytchien@mit.edu}
\affiliation{Center for Theoretical Physics, Massachusetts Institute of Technology, Cambridge, MA 02139, USA}

\author{Daekyoung Kang}
\email[]{dkang@fudan.edu.cn}
\affiliation{Key Laboratory of Nuclear Physics and Ion-beam Application (MOE) and Institute of Modern Physics, Fudan University, Shanghai, China 200433}

\author{Kyle Lee}
\email[]{kunsu.lee@stonybrook.edu}
\affiliation{C.N. Yang Institute for Theoretical Physics, Stony Brook University, Stony Brook, NY 11794,
  USA}
 \affiliation{Department of Physics and Astronomy, Stony Brook University, Stony Brook, NY 11794, USA}

\author{Yiannis Makris}
\email[]{yiannis@lanl.gov}
\affiliation{Theoretical Division T-2, Los Alamos National Laboratory, Los Alamos, NM, 87545, USA}

\begin{abstract}
We introduce a new approach for jet physics studies using subtracted cumulants of jet substructure observables,   
which are shown to be insensitive to contributions from soft-particle emissions uncorrelated with the hard process.
Therefore subtracted cumulants allow comparisons between theoretical calculations and experimental measurements without the complication of large background contaminations such as underlying and pileup events in hadron collisions. 
We test our method using subtracted jet mass cumulants by comparing Monte Carlo simulations to analytic calculations performed using soft-collinear effective theory. We find that, for proton-proton collisions, the method efficiently eliminates contributions from multipart on interactions and pileup events. We also find within theoretical uncertainty that our analytic calculations are in good agreement with the subtracted cumulants calculated by using ATLAS jet mass measurements.
\end{abstract}

\preprint{LA-UR-18-31092}
\preprint{MIT-CTP 5088}

\maketitle

\section{Introduction}

Jets have become essential objects of study at high energy colliders. They are produced ubiquitously in hard scattering processes as well as hadronic decays of heavy particles. Tremendous progress has been made to understand and use jets for testing the standard model and searching for new physics, with reliable Monte Carlo simulations \cite{Sjostrand:2007gs,Bellm:2015jjp} of high-energy collisions as well as useful jet substructure analysis tools \cite{Larkoski:2017jix,Andrews:2018jcm}. However, theoretical precision is often limited by the need to model soft radiation such as from multiparton interactions (MPI) in hadron collisions, which are insensitive to the hard process in a wide range of energies. Other contributions to the underlying event (UE) are either calculable (e.g. initial state radiation (ISR)) or relatively small (e.g. hadronization).  In high-luminosity (HL) collisions there can also be a large number of uncorrelated pileup (PU) events producing a significant background of soft particles. With accurate vertex determination using charged-particle tracks, one can remove PU charged particles but still not PU neutral particles. Also, heavy-ion collisions (HIC) can produce a large number of soft particles ($\sim \mathcal{O}(10^5)$) through the interactions among a large number of nucleons in the nuclei. The soft particles are observed to be mostly uncorrelated with the hard process of interest and exhibit novel collective behaviors \cite{Voloshin:1994mz,Poskanzer:1998yz}. However, correlated effects certainly exist and significantly quench the jets \cite{Bjorken:1982tu,Connors:2017ptx}.

It is clear that jet observables are affected dramatically by an uncorrelated large background, which can overwhelm the correlated effects one wants to probe such as jet modifications and medium responses in HIC. Many subtraction techniques \cite{Cacciari:2007fd,Feige:2012vc, Soyez:2012hv, Larkoski:2014bia,Krohn:2013lba,Cacciari:2014gra,Bertolini:2014bba,Berta:2014eza,Larkoski:2014wba,Komiske:2017ubm,Soyez:2018opl} have been developed in order to correct jets back to their true compositions. Ideally one would like the subtraction to work for each jet, which is impossible due to the intrinsic ambiguity between signal and background particles. One could then hope to remove the background and correct jet observable distributions statistically. The precision of the background subtraction will then have to be quantified when comparing measurements to analytic calculations.

In this paper, instead of relying on algorithms to remove soft particles out of jets, we provide an alternative approach for comparing theoretical calculations directly to experimental measurements without the complication of modeling soft uncorrelated emissions (SUEs). Specifically, we define subtracted cumulants which cancel the contributions from SUEs. This approach was first introduced in the context of the transverse energy of Drell-Yan processes \cite{Kang:2018agv}. Here we extend its use to jet substructure observables to which SUEs additively contribute. Additive contributions from uncorrelated emissions can be easily removed, and the resulting subtracted cumulant is thus useful for precision jet physics studies. The jet mass, $m_J$ is a classic observable which receives additive contributions from SUEs. In this paper, as a proof of concept, we focus only on the first cumulant of jet invariant mass in proton-proton collisions. It is straightforward to extend our study to other jet substructure observables and higher cumulants.

The rest of the paper is organized as follows. We first give the definition of subtracted jet mass cumulant $\Delta$, and we demonstrate its robustness against SUEs using \textsc{Pythia} Monte Carlo simulations. Since jet substructure observables highly depend on the jet-initiating partons we also show the sensitivity of $\Delta$ to quark-gluon jet fractions. Finally, we show that the comparisons of our theoretical predictions performed using soft-collinear effective theory (SCET) \cite{Bauer:2000ew,Bauer:2000yr,Bauer:2001ct,Bauer:2001yt,Bauer:2002nz,Beneke:2002ph} to the results computed from ATLAS measurements are in a good agreement.

\section{Definition of the observable}

For a jet substructure observable, $e$, which receives additive contributions from individual particles within a jet, we have
\begin{equation}
  e = \sum_{i \in \text{jet}} \hat{e}^{(i)}\;,
\end{equation}
where $\hat{e}^{(i)}$ depends on the four-momenta of the $i$-th particle in the jet. In the presence of SUEs, because of the additivity, the observable can be decomposed into two terms,
\begin{equation}\label{eq:l-deco}
  e = \sum_{i \in \text{signal}} \hat{e}^{(i)} + \sum_{i \in \text{SUEs}} \hat{e}^{(i)} \equiv \es+\eb \;,
\end{equation}
where \textquotedblleft S\textquotedblright{} refers to the signal contributions which are correlated with the hard process and \textquotedblleft B\textquotedblright{} to the background from SUEs. Here $\eb$ is the background contribution statistically independent of $\es$. Its probability density does not depend on the kinematics and details of the hard process, such as the jet energy, angular direction, and the flavor of the initiating parton. Let $P_\text{S}(\es)$, $P_\text{B}(\eb)$, and $P(e)$ denote probability densities of the observables $\es$, $\eb$, and $e$, respectively. Since SUEs are uncorrelated with the signal, the probability density at the values $\es$ and $\eb$ is simply a product of uncorrelated distributions $P(\es,\eb) = P_\text{S}(\es)P_\text{B}(\eb)$. Then, $P(e)$ is given by
\begin{eqnarray}\label{eq:conv.}
  P(e) &=& \int d\es \; d\eb \;\delta (e-\es-\eb) P(\es,  \eb )
  \nn\\
  &=& \int d\eb \; P_\text{S}(e-\eb) P_\text{B}(\eb )\;,
\end{eqnarray}
which has a convolution form. The cumulants $\kappa_n(e)$ are defined using the cumulant-generating function $K(t)$,
\begin{equation}
  \label{eq:cumu}
    K(t)=\sum_{\rm n=1}^\infty \kappa_n\frac{t^n}{n!}=\log \langle {\rm exp}(t e)\rangle\;,
\end{equation}
where $\langle\cdots\rangle$ denotes the expectation value. Note the additivity of cumulants: $\kappa_n(e)=\kappa_n(\es)+\kappa_n(\eb)$ which will allow us to cancel uncorrelated contributions in the subtraction between cumulants. Also, cumulants are in one-to-one correspondence with moments $\langle e^n\rangle$: $\kappa_1(e)=\langle e\rangle$, $\kappa_2(e)=\langle e^2\rangle-\langle e\rangle^2$, $\kappa_3(e)=\langle e^3\rangle-3\langle e^2\rangle\langle e\rangle+2 \langle e\rangle^3$, etc.

Although Eq.(\ref{eq:l-deco}) is by definition true for all additive observables, jet substructure observables defined with the standard jet axis determined via E-scheme is subject to an axis recoil effect due to soft radiation, which correlates the soft background and signal radiation. However, such effects are shown to be power-suppressed \cite{Ellis:2010rwa} for many of the additive observables, including the jet mass we study here. Also, as we show later using simulation data, this effect is expected to be small for subtracted cumulants due to the uniformity of soft radiation. Alternatively one may use a recoil free axis such as the winner-take-all axis~\cite{Bertolini:2013iqa}. There are also many additive jet observables which are insensitive to jet axis direction which would not be subject to such recoil effects~\cite{Banfi:2004yd}.

We define the jet substructure observable $\hat{\tau}$ which is closely related to the jet invariant mass and receives additive contributions from signal and background,
\begin{equation}
  \label{eq:hat-tau}
\hat{\tau} = 2 \cosh(\eta) \sum_{i\in \text{jet}} p_{i}^{+} = \hat{\tau}_S +\hat{\tau}_B  = \frac{m_J^2}{p_T} \left[ 1+ \mathcal{O}\lp\frac{m_J^2}{p_T^2}\rp \right]\;,
\end{equation}
where $p_T$ and $\eta$ are the jet transverse momentum and pseudorapidity with respect to the beam axis, and $p^{+}=p^0-\vec{n}\cdot \vec{p}$ is the small light-cone component of the constituent's momentum with respect to the jet axis $\vec{n}$. Then for the dimensionless observable $\tau = \hat{\tau}/p_T$, up to corrections suppressed by $(p^{\rm SUE}_TR^2/p_T)^2$, where $R$ is the jet radius, Eq.(\ref{eq:conv.}) becomes
\begin{equation}
  \label{eq:convolution}
  \frac{d\sigma}{d\tau dp_Td\eta}(\tau,p_T) = \int d\hat{\tau}_{\text{B}} \frac{d\corr{\sigma}}{d\tau dp_Td\eta} \lp \tau- \frac{ \hat{\tau}_{\text{B}}}{p_T}, p_T \rp f(\hat{\tau}_{\text{B}})\;,
\end{equation}
where the function $f$ is the normalized probability distribution of the SUE contribution to the observable $\hat{\tau}$. It was shown that this convolutional expression with a simple model for $f$ well describes the MPI contribution in Monte Carlo simulations \cite{Stewart:2014nna,Hoang:2017kmk,Kang:2018agv} and experimental measurements \cite{Kang:2018jwa}.
Note that the expression in \eq{convolution} resembles the factorization of hadronization contributions derived using operator product expansion \cite{Lee:2006nr}, and hadronization is correlated with the hard process. Here, $f$ only includes SUEs and is observed to be independent of the jet pseudorapidity in the plateau region. Due to the similar convolution structure for hadronization effects, hadronization effects are also largely removed in the subtracted cumulant we define below for proton-proton collisions.

For the first $\tau$ cumulant (the first moment, equivalently), which is denoted by $\langle \tau \rangle$ and is a function of jet $p_T$ and $\eta$,
\begin{equation}\label{eq:taupT}
\langle \tau \rangle
=\left(\frac{d\sigma}{dp_Td\eta}\right)^{-1}  \int d\tau \, \tau \frac{d\sigma}{d\tau dp_Td\eta }
=\langle \corr{\tau} \rangle +\frac{\Omega_f}{p_T}
\,,\end{equation}
where $\langle \corr{\tau} \rangle$ is the first cumulant in the absence of SUEs and $\Omega_f = \int d\hat{\tau}_{\text{B}}\, \hat{\tau}_{\text{B}} f(\hat{\tau}_{\text{B}})$ is independent of hard scale $p_T$. Therefore one can define SUE-independent observable by taking the derivative of the $p_T$-weighted cumulant:
\begin{equation}
\frac{d}{dp_T}p_T \langle \tau \rangle
=\frac{d}{dp_T}p_T \langle\corr{\tau}  \rangle\;.
\,\end{equation}
For a binned cross section $\sigma^{[i,j]}$ of the $i$-th bin in $\tau$ and $j$-th bin in jet $p_T$, the $\tau$ cumulant of the $j$-th $p_T$ bin is the following:
\begin{equation}\label{eq:tauj}
\langle \tau \rangle^{[j]}
= \frac{ \sum_i \tau^{[i]} \sigma^{[i,j]}  }{\sum_i\sigma^{[i,j]}}
=\langle \corr{\tau} \rangle^{[j]} +  \Omega_f \langle p_{T}^{-1} \rangle^{[j]}
\,,\end{equation}
where $\tau^{[i]}$ is the central value of the $i$-th $\tau$ bin and $\langle p_{T}^{-1} \rangle^{[j]} =  (\int_{{\rm bin}~j} p_T^{-1}d\sigma  )/\sum_i \sigma^{[i,j]}$. We then define the subtracted cumulant with the same mass dimension as $\tau$,
\begin{equation}\label{eq:delta-tau}
\Delta_\tau^{jk}=
\langle \tau \rangle^{[j]}
-\langle \tau \rangle^{[k]}\frac{\langle p_{T}^{-1} \rangle^{[j]}}{\langle p_{T}^{-1} \rangle^{[k]}}
=
\langle \corr{\tau} \rangle^{[j]}
-\langle \corr{\tau} \rangle^{[k]}\frac{\langle p_{T}^{-1} \rangle^{[j]}}{\langle p_{T}^{-1} \rangle^{[k]}}
\,.\end{equation}
The model function dependence vanishes and we are left with purely signal-correlated contributions. Note that we do not have to assume any specific form for the model function, $f(\hat{\tau}_{\text{B}})$.

\begin{figure}[t!]
  \centerline{\includegraphics[width = 0.48 \textwidth]{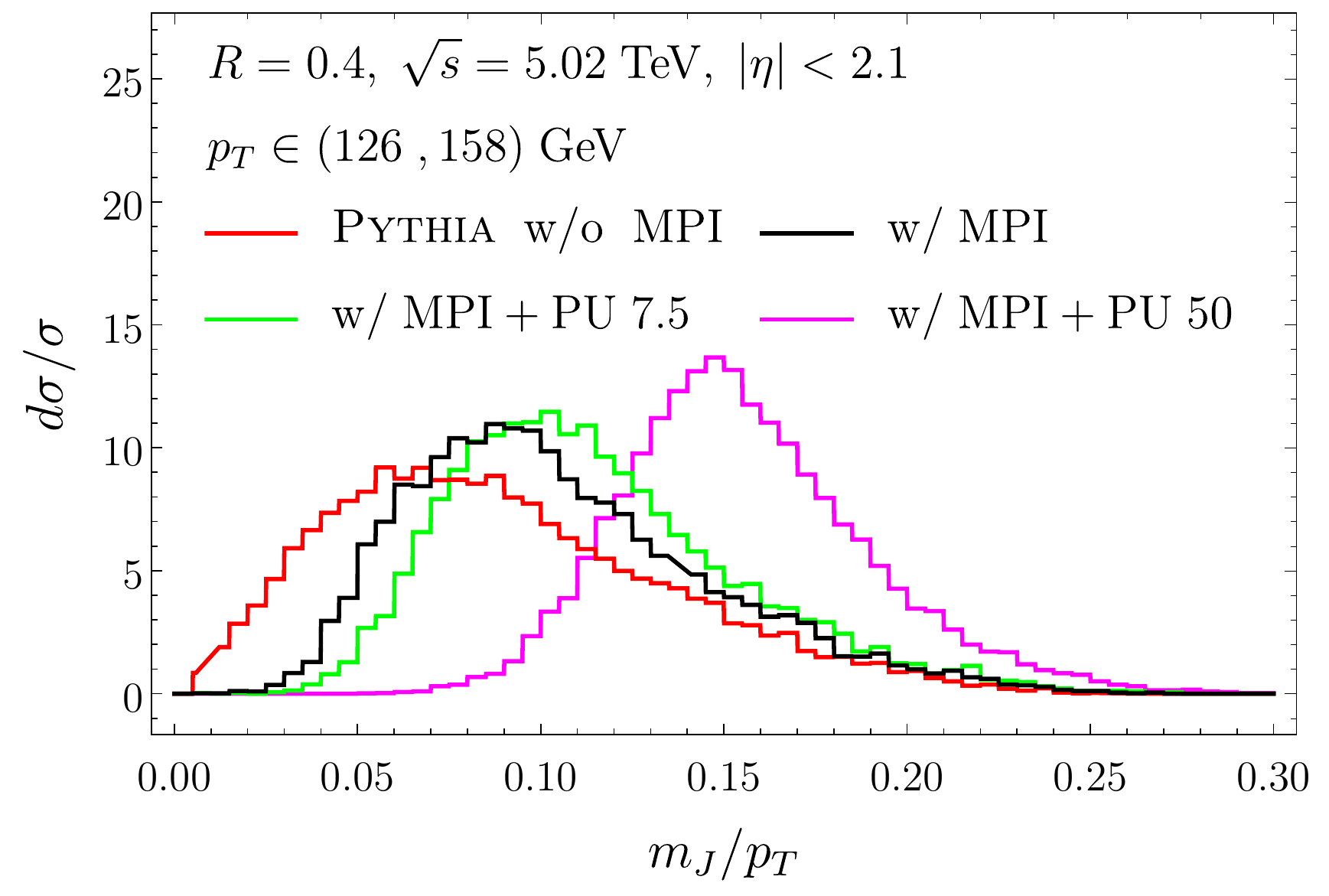}}
  \caption{Jet invariant mass distributions simulated in \textsc{Pythia} at parton level (red), hadron level with underlying events (black), pileup with $\langle N_{\text{PU}} \rangle = 7.5$ (green) and 50 (magenta). }
  \label{fig:diff}
\end{figure}

\section{Removal of soft uncorrelated emissions}

We discuss and demonstrate using \textsc{Pythia} simulations that subtracted cumulants are indeed insensitive to MPI and PU contributions in proton-proton collisions. We compare the results with the perturbative calculation performed in \cite{Kang:2018jwa} at next-to-leading logarithmic and next-to-leading order accuracy (\NLLp+NLO) using SCET. (See also \cite{Dasgupta:2012hg,Chien:2012ur,Jouttenus:2013hs,Liu:2014oog,Hornig:2017pud} for previous jet mass calculations.) Within theoretical uncertainties, the calculation agrees well with the simulations even for a large number of PU events.

The \NLLp +NLO result is obtained by matching the resummed and fixed order results,
\begin{equation}
  \frac{d\sigma^{\text{\NLLp +NLO}}}{d\tau dp_Td\eta} =  \frac{d\sigma^{\text{\NLLp}}}{d\tau dp_Td\eta} +  \frac{d\sigma^{\text{NLO}}}{d\tau dp_Td\eta} - \frac{d\sigma^{\text{NLO-sing.}}}{d\tau dp_Td\eta}
\,,\end{equation}
where $d\sigma^{\text{\NLLp}}$, $d\sigma^{\text{NLO}}$, and $d\sigma^{\text{NLO-sing.}}$ are the resummed, fixed-order, and fixed-order singular cross sections, respectively. The NLO result\footnote{
In this paper, `NLO' means the fixed-order calculation at $\mathcal{O}(\alpha_s)$ accuracy. Since $\mathcal{O}(\alpha_s)$ is the first order at which the mass is nonzero, $\mathcal{O}(\alpha_s)$ is sometimes referred to as the `LO' contribution \cite{Salam:2009jx}.
}is obtained using MadGraph 5~\cite{Alwall:2014hca}. For the simulation, we use \textsc{Pythia} 8~\cite{Sjostrand:2006za,Sjostrand:2007gs} with the ATLAS-A14-variation-2$+$ tune.  We study the effect of MPI on subtracted cumulants by switching on and off its contribution in \textsc{Pythia}. PU events are simulated by soft QCD processes and added on top of signal events, and the PU event number follows a Poisson distribution with the mean $\langle N_{\rm PU}\rangle$. Here we present results for $\langle N_{\rm PU}\rangle = $ 7.5 and 50. Jets are reconstructed using the anti-$k_t$ algorithm \cite{Cacciari:2008gp} implemented in \textsc{FastJet}~\cite{Cacciari:2011ma}.
 
\begin{figure}[t!]
  \centerline{\includegraphics[width = 0.48 \textwidth]{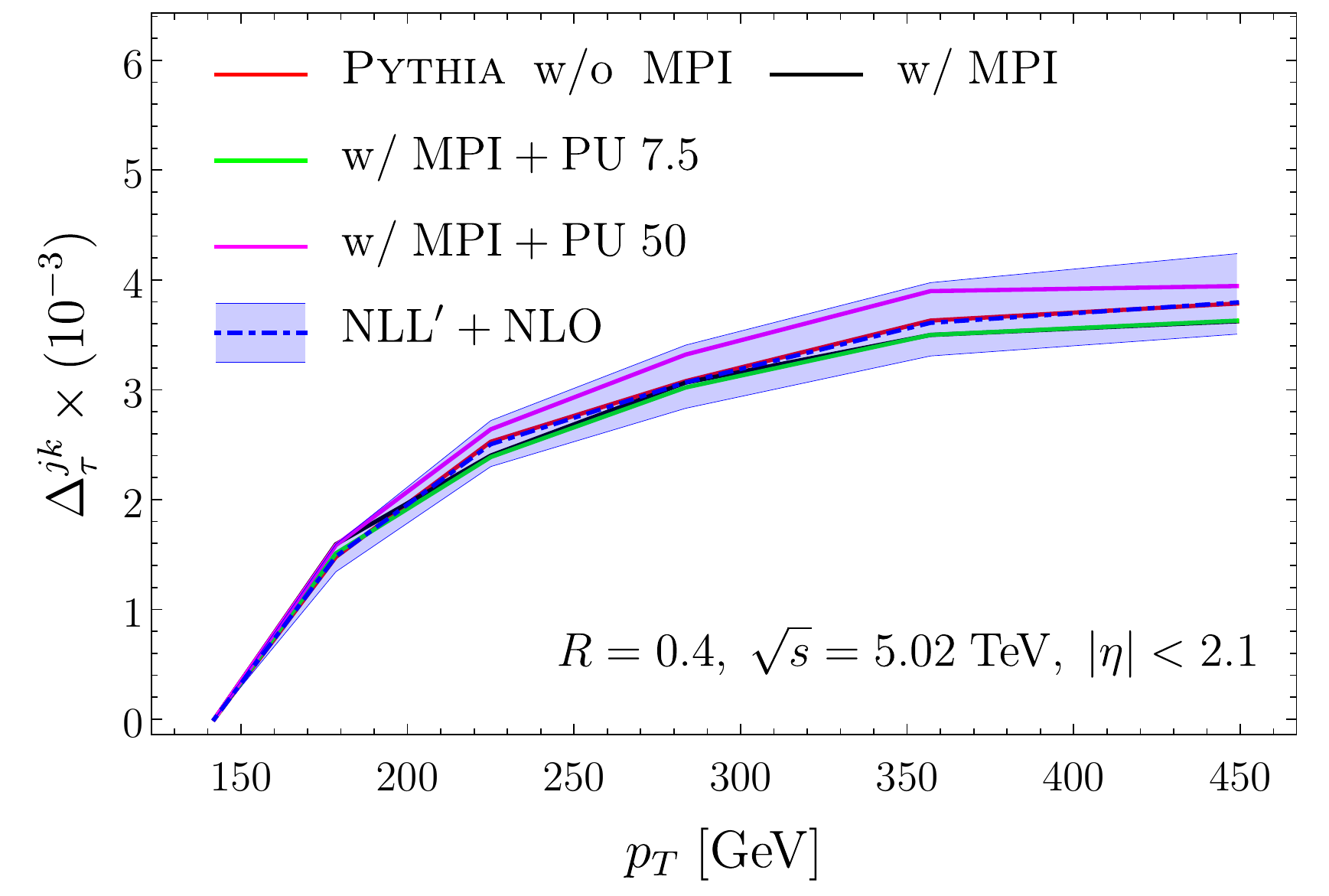}}
  \caption{Subtracted jet mass cumulants $\Delta_\tau^{jk}$ from perturbative calculation (blue band) and \textsc{Pythia} simulations without MPI (red), as well as with MPI (black) and pileup with $\langle N_{\text{PU}} \rangle = 7.5$ (green) and 50 (magenta). The following transverse momentum bins are used:
  $p_T \in\left\{[126,158], [158,199], [199,258], [258,316], [316,398], [398,500]\right\}$.}
  \label{fig:pileup}
  \vspace{- 0.1 cm}
\end{figure}

We first show in FIG.~\ref{fig:diff} the contributions of MPI and PU to the $m_J/p_T$ distributions. Both MPI and PU affect the peak position of the distribution significantly, especially for lower $p_T$ jets with a large $\langle N_{\rm PU}\rangle$. FIG.~\ref{fig:pileup} then shows the results of subtracted jet mass cumulants. The blue band is the theoretical uncertainty of the \NLLp+NLO calculation estimated by varying characteristic energy scales with a factor of two. Remarkably, the simulation results from \textsc{Pythia} for different cases with and without MPI or PU contributions all agree with the analytic calculation of the signal distribution within theoretical uncertainty. This clearly demonstrates that the proposed subtracted cumulants largely mitigate contributions from UE and PU.

\section{Modification for high luminosity collisions}
\label{sec:HLC}

For the situation with large background contamination from PU at HL-LHC or UE in HIC, the jet $p_T$ is significantly altered by SUEs and the jet mass is no longer an additive observable from jet constituents. Therefore, we instead consider the observable, $\hat{\tau}$, defined in Eq.(\ref{eq:hat-tau}) which is explicitly additive. Note that the jet direction $\vec{n}$ is assumed to be only mildly affected by a large but approximately uniform background, or one can use a recoil-free axis \cite{Bertolini:2013iqa}. On the other hand, since SUE contamination alters the value of jet $p_T$ significantly, in order to compare subtracted cumulants between experiment and theory we need to correct for the jet $p_T$ bin migration. This can be effectively achieved using the area subtraction method \cite{Cacciari:2007fd,Cacciari:2009dp}, and we refer to the corrected $p_T$ as $\hat{p}_{T}$. The subtracted cumulants for $\hat{\tau}$ are defined as follows,
\begin{equation}\label{eq:delta-tauhat}
\Delta^{jk}_{\hat{\tau}}=
\langle \hat{\tau}\rangle^{[j]}
-\langle \hat{\tau} \rangle^{[k]}
=
\langle \corr{\hat{\tau}} \rangle^{[j]}
-\langle \corr{\hat{\tau}} \rangle^{[k]}
\,,\end{equation}
where the indices $j,k$ label the $\hat{p}_T$ bins.
Note that the subtracted cumulant of $\hat\tau$ above is different from \eq{delta-tau} in $p_T$-weighting factor.

\begin{figure}[t!]
\centerline{\includegraphics[width = 0.5 \textwidth]{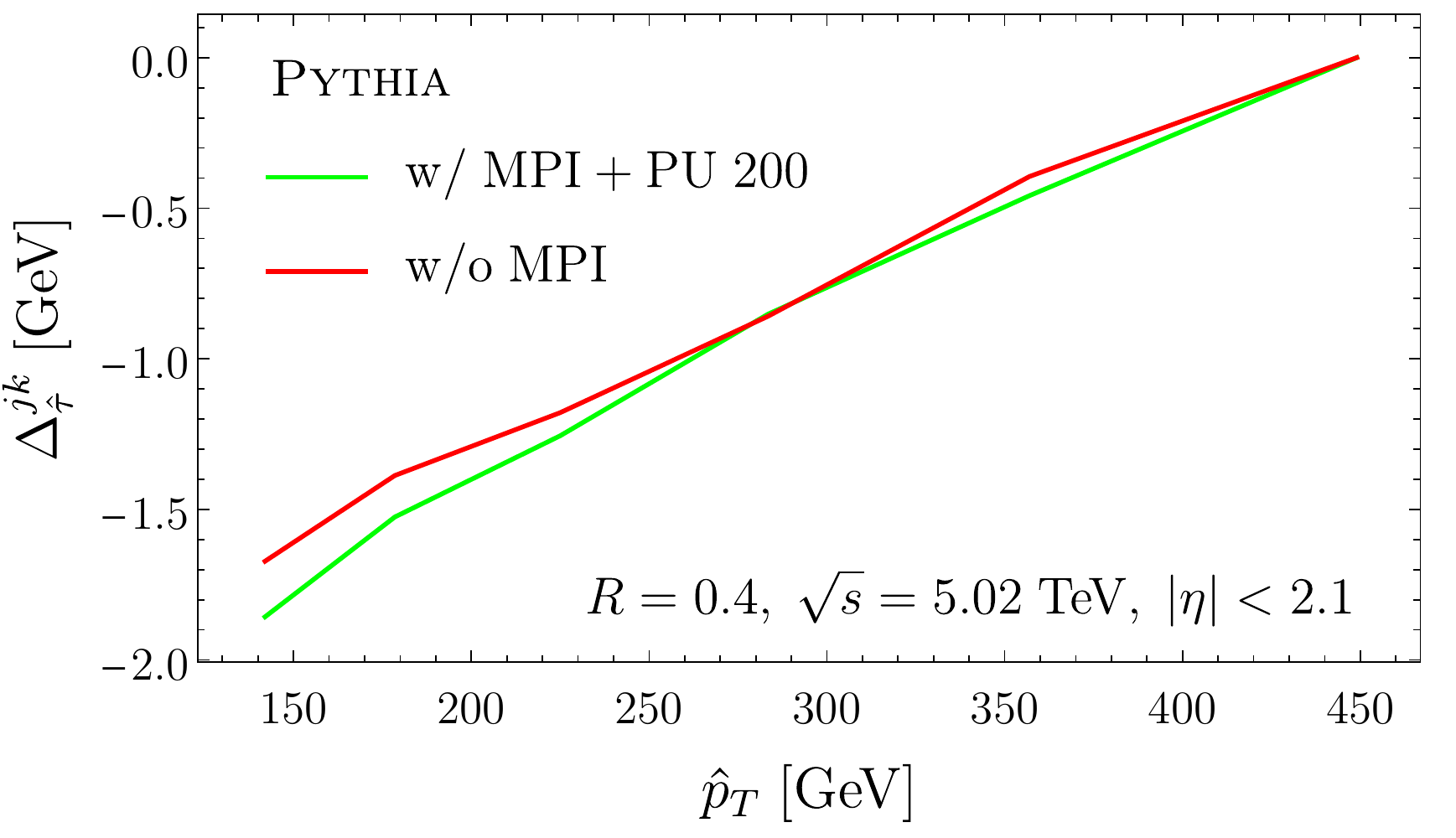}}
  \caption{Subtracted cumulants $\Delta_{\hat{\tau}}$ obtained using \textsc{Pythia} simulations at the parton level (red) and hadron level with large PU contaminations of $ \langle N_{\text{PU}} \rangle  = 200$ (green). }
  \label{fig:dm2}
\end{figure}

In FIG.~\ref{fig:dm2} we demonstrate the robustness of $\Delta^{jk}_{\hat{\tau}}$ against large SUEs by comparing the \textsc{Pythia} partonic result to the one including MPI and PU with $\langle N_{\text{PU}} \rangle = 200$, which is typical at HL-LHC and can give an indication of how this observable removes SUEs in HIC. Note the remarkable agreement between the two results. In practice, we use the approximation $\hat{\tau} \simeq m_J^2/p_T$ which is in terms of the well-studied invariant mass. For this reason and in contrast to the previous plots, we choose to subtract the highest, instead of the lowest, $\hat{p}_T$ bin where this approximation is more accurate.

\section{Sensitivity to quark/gluon jet fraction}

We discuss the sensitivity of subtracted cumulants to quark and gluon jet fractions, $f_q = 1 - f_g$ and $f_g$, respectively. Assuming that the fractions vary slowly within each $p_T$ bin $j$, the $\tau$ distribution is a weighted sum of the corresponding quark and gluon distributions,
\begin{equation}
  \frac{d\sigma^{[j]}}{d\tau} = f_{g}^{[j]} \; \frac{d\sigma^{[j]}_{g}}{d\tau}+ (1-f_{g}^{[j]} )\;  \frac{d\sigma^{[j]}_{q}}{d\tau},
\end{equation}
and similarly for $\langle \tau\rangle$,
\begin{equation}
  \langle\tau\rangle^{[j]} = f_{g}^{[j]} \; \langle\tau\rangle^{[j]}_{g}+ (1-f_{g}^{[j]} )\;  \langle\tau\rangle^{[j]}_{q},
\end{equation}
Since $0< f_g <1$ and $ \langle\tau\rangle^{[j]}_{g} > \langle\tau\rangle^{[j]}_{q}$, we have
$  \langle\tau\rangle^{[j]}_{q}  < \langle\tau\rangle^{[j]} < \langle\tau\rangle^{[j]}_{g}$.
The subtracted cumulants are
\begin{eqnarray}\label{eq:tau-gfrac}
  \Delta_\tau^{jk} &=&  \Delta^{jk}_{\tau,q} + \lb f_g^{[j]} (\langle\tau\rangle_{g} - \langle\tau\rangle_{q})^{[j]}
 \nn \\ && \qquad  \qquad
  - f_g^{[k]} (\langle\tau\rangle_{g} - \langle\tau\rangle_{q})^{[k]} \frac{\langle p_{T}^{-1} \rangle^{[j]}}{\langle p_{T}^{-1} \rangle^{[k]}}\rb\;.
\label{eq:delta-gfrac}
\end{eqnarray}
We use \textsc{Pythia} to simulate pure quark and gluon jets, and we mix the samples manually using the parametrized function $f_g(p_T;a,b)$ (see Appendix for details). Within the $p_T$ range of interest we examine two scenarios in which the gluon jet fraction is larger (model 1) or smaller (model 2) than the expected value in $pp$ collisions.

\begin{figure}[t!]
  \centerline{\includegraphics[width = 0.48 \textwidth]{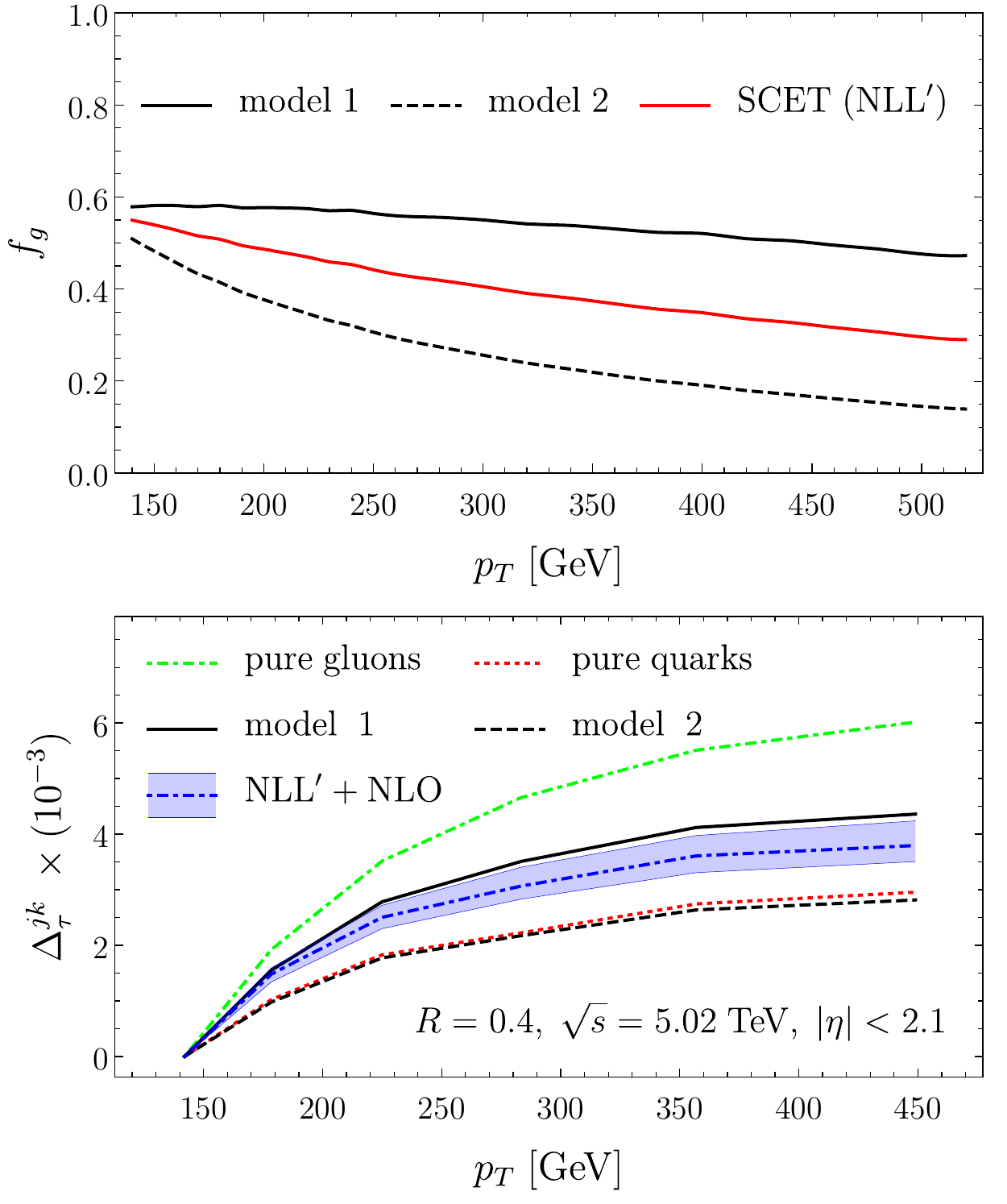}}
  \caption{Top: Gluon jet fractions for the two models in this analysis (see text and Appendix~\ref{sec:fractions}). Bottom: Results of subtracted cumulants from analytic calculation (blue band) and \textsc{Pythia} simulations with the gluon fractions from models 1 and 2, as well as pure quark and gluon jets.}
  \label{fig:AA}
  \vspace{- 0.1 cm}
\end{figure}

FIG.~\ref{fig:AA} shows the gluon jet fraction and subtracted cumulant as a function of jet $p_T$ for model 1 and model 2, as well as theoretical predictions at \NLLp{} accuracy for $pp$ collisions. We find that a change of quark-gluon jet fraction can induce a significant change of the subtracted cumulant distinguishable with the theoretical precision. Precise measurements of subtracted cumulants of inclusive jets (gluon-enriched) and photon-tagged jets (quark-enriched) will then give useful information about the different quark-gluon jet fractions as well as subtracted cumulants of pure quark and gluon jet samples. Since quark and gluon jets are initiated by partons with different color charges, one expects that the two are quenched differently and thus their fractions may change from proton-proton to HIC \cite{Spousta:2015fca,Chien:2015hda}. The fraction change can induce modifications of jet substructure which should be disentangled from the jet-by-jet modification, for which subtracted cumulants can be very useful.

\begin{figure}[t!]
  \centerline{\includegraphics[width = 0.48 \textwidth]{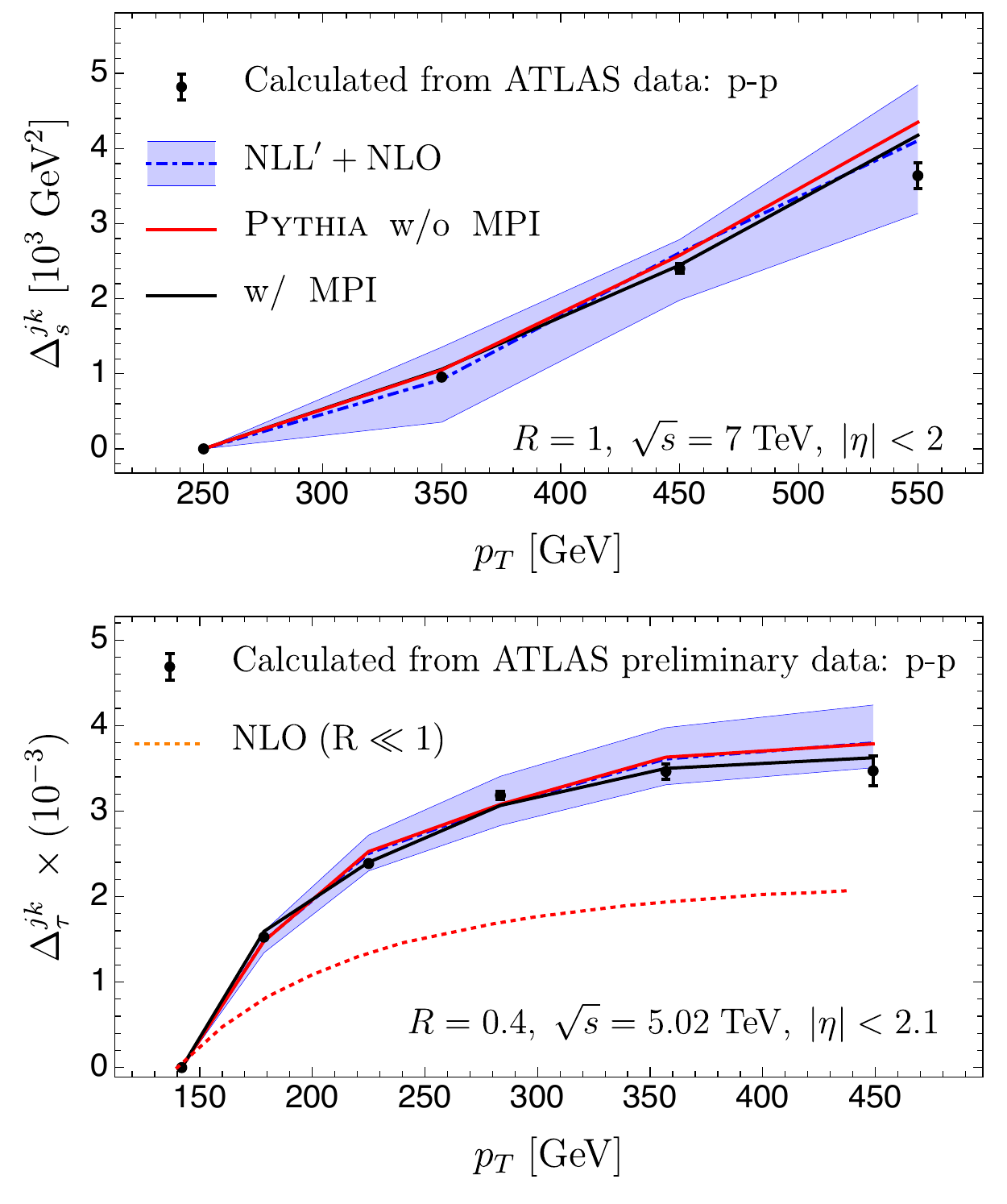}}
  \caption{Comparison of the subtracted cumulant results from \NLLp+NLO calculation (blue band) with \textsc{Pythia} simulations with (black) and without (red) MPI and hadronization, as well as subtracted cumulants calculated from the experimental data measured by the ATLAS~\cite{ATLAS:2012am, ATLAS:2018jsv}. The top and bottom panels correspond to the collisional center of mass energy at $\sqrt{s} = 7$ TeV and$\sqrt{s}=5.02$ TeV, respectively.}
  \label{fig:sub}
  \vspace{- 0.1 cm}
\end{figure}

\section{Comparison with experimental data}

We compare our analytic calculation and simulation to subtracted cumulants calculated from the experimental data measured by the ATLAS Collaboration at the LHC with the collisional center of mass energies 7 TeV \cite{ATLAS:2012am} and 5.02 TeV \cite{ATLAS:2018jsv}.

FIG.~\ref{fig:sub} shows the results for the \NLLp+NLO calculation (blue band) and \textsc{Pythia} simulations with (black) or without (red) MPI effect and hadronization. The data points are calculated from ATLAS measurements of jet mass distributions. The error bars include only the statistical uncertainty and are calculated from the variance of  $\langle\tau\rangle^{[j]}$: $\sqrt{ \text{Var}[\tau]^{[j]}/N^{[j]}}$, where $\text{Var}[\tau]^{[j]}$ is the variance of the $\tau$ distribution and $N^{[j]}$ is the total number of jets estimated from the integrated luminosity: $\mathcal{L}_{\text{int.}}\times d\sigma^{[j]}$. The statistical error in these experiments is small resulting in the small error bars in the plots. Including the systematic uncertainty requires experiment details and is beyond the scope of this work. For the 7 TeV case, only the differential distributions in jet mass are available rather than $\tau=m_J^2/p_T^2$ thus we redefine $\Delta$ in terms of the cumulant of $s=m_J^2$ as follows,
\begin{equation}
\Delta^{jk}_s=
\langle s\rangle^{[j]}
-\langle s \rangle^{[k]}\frac{\langle p_{T} \rangle^{[j]}}{\langle p_{T} \rangle^{[k]}}
=
\langle \corr{s} \rangle^{[j]}
-\langle \corr{s} \rangle^{[k]}\frac{\langle p_{T} \rangle^{[j]}}{\langle p_{T} \rangle^{[k]}}
\,.\end{equation}
 This redefinition is only necessary due to the large $p_T$ bin sizes in the experiment. The average values $\langle p_T \rangle^{[j]}$ are not given in \cite{ATLAS:2012am} and we use the ones generated by \textsc{Pythia} including hadronization and underlying event contributions since these quantities are well described by simulations. For both the 7 and 5.02 TeV cases, we find that the results of analytic calculations and simulations are in good agreement with the experimental data.

A fixed-order expression for mean square jet mass  $\langle m^2 \rangle$ for quark or gluon jets in small radius limit ($R\ll 1$) is given in Ref.~\cite{Salam:2009jx}: 
$\langle \tau \rangle_i \simeq \langle m^2 \rangle_i/p_T^2 \simeq C_ i\frac{\alpha_s}{\pi} R^2
\,,$
where $i=q,g$ and $C_q=\tfrac{3}{8}C_F$ and $C_g=\tfrac{7}{20}C_A+\tfrac{1}{20}n_f T_R$. 
By inserting it into \eq{tau-gfrac}, the fixed-order subtracted cumulant is given by
\begin{eqnarray}
\label{eq:approx}
\Delta^{jk}_\tau\simeq\frac{\alpha_s}{\pi}R^2 
	&& \left[ C_q\frac{p_T^{[j]}-p_T^{[k]}}{p_T^{[j]}} \right .
\nn\\&& \quad
\left. +(C_g-C_q)\frac{ p_T^{[j]}\, f_g^{[j]}-p_T^{[k]}\,f_g^{[k]}}{p_T^{[j]}} \right]
\,,
\end{eqnarray}
We apply \eq{approx} to the case $R=0.4$ in the lower panel of FIG.~\ref{fig:sub} and we obtained  the NLO curve (red dotted)
 by using the gluon fraction in FIG.~\ref{fig:AA}. The result underestimates the data and the calculation including resummation and this suggests that resummation plays an important role. This can be understood from the fact that the jet mass has a maximum allowed value $m_J^\text{max} \simeq p_T^J \tan(R/2)\simeq 0.2\, p_T^J$ which is not sufficiently far from the location of the peak in FIG.~\ref{fig:diff} where resummation is needed. 
 The cumulants of $e^+e^-$ thrust \cite{Abbate:2012jh} also show a significant difference between the fixed-order result at $\mathcal{O}(\alpha_s)$ and the resummed result at NLL.

\section{Conclusions}

\begin{figure*}[t!]
  \centerline{\includegraphics[width = 0.98 \textwidth]{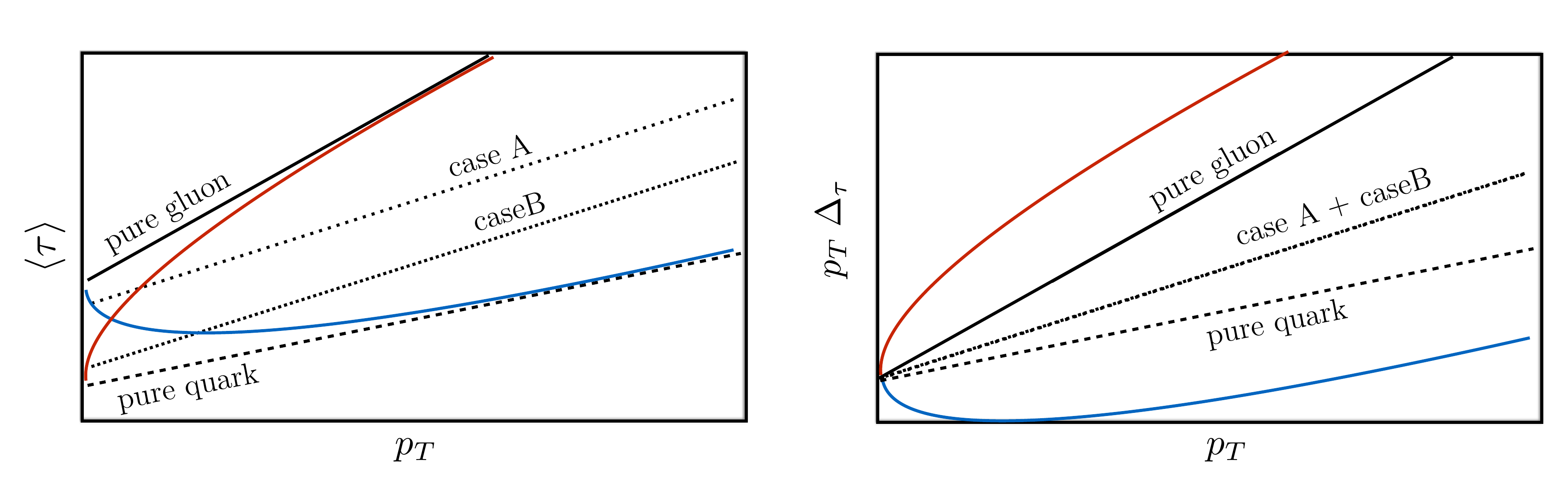}}
  \caption{Cumulants (left panel) and subtracted cumulants (right panel) corresponding to different quark and gluon jet fractions: pure quark (dashed), pure gluon (solid) and two interpolations between pure quark or gluon across jet $p_T$ (red and blue), as well as generic cases A and B (dotted) which lie between the pure quark and gluon cases.}
  \label{fig:appA}
\end{figure*}

In this paper, we extend the work in \cite{Kang:2018agv} to jet substructure observables and introduce the new method of comparing theoretical calculations of jet substructure observables to data using subtracted cumulants. An advantage of our observable, compared to event by event subtraction techniques, is that our approach can be applied to existing data for which only the jet four-momenta are available. Furthermore it does not require any tunes or Monte Carlo input since it does not depend on any parameters or models.  
The method makes the comparison insensitive to soft uncorrelated emissions such as multiparton interactions and pileup without using background subtraction algorithms to correct each jet or having to model uncorrelated effects. Our theoretical prediction at \NLLp+NLO accuracy using SCET shows an excellent agreement with the subtracted cumulants calculated from two independent ATLAS jet mass measurements and those from \textsc{Pythia} simulations. We also demonstrate that subtracted jet substructure cumulants remove large background contaminations up to 200 pileup events. Its robustness makes subtracted cumulants useful for jet studies at the high-luminosity LHC and in the heavy-ion collisions, where the identification of signal jets is challenged by a large background. We also show that subtracted cumulants are sensitive to the change of quark-gluon jet fraction. This could allow for precise determination of the fraction and its modification in heavy-ion collisions, which will be useful for discriminating possible medium effects and contributions. For example, in addition to the p-p jet mass measurements, in Ref.~\cite{ATLAS:2018jsv}, exist also preliminary Pb-Pb data for the same transverse momentum. It will  be very interesting to see a comprehensive analysis and a comparison of the subtracted jet mass cumulants from the two measurements and what that reveals for the medium induced effects. The mitigation of UE with flow modulation in HIC will be studied in future work.

\begin{acknowledgments}
The authors would like to thank Yongsun Kim, Yen-Jie Lee, Christopher Lee, Duff Neill, Felix Ringer, Iain Stewart, Jesse Thaler and Ivan Vitev for useful conversations during the completion of this work. Y.T.C. is supported by the LHC Theory Initiative Postdoctoral Fellowship under the National Science Foundation grant PHY-1419008. D.K. is supported by the National Natural Science Foundation of China under Grant No.~11875112.
K.L. is supported by the National Science Foundation under Grants No.~PHY-1316617 and No.~PHY-1620628. Y.M. is supported by the DOE Office of Science under Contract No.~DE-AC52-06NA25396, the Early Career Program (Christopher Lee, P.I.) and the LDRD Program at LANL.
\end{acknowledgments}


\appendix
    \section{Dependence of subtracted cumulants on quark and gluon jet fraction}
\label{sec:fractions}

We give the details of the parametrization used in FIG.~\ref{fig:AA} for the two models with different quark and gluon jet fractions. The gluon fraction, $f_g$, has the following power-law modification form:
\begin{equation}
  f_g(p_T;a,b)  = f^{\text{NLL}}_g(p_T) \lp \frac{p_T}{a}  \rp^{b}\;,
\end{equation}
where $a$ and $b$ can be varied. The function $f_{g}^{\text{NLL}}$ is the analytic result extracted from the \NLLp\, calculation of the inclusive cross section \cite{Kang:2016mcy} and tt is the gluon fraction at the hard scale $ \mu_H = p_T$. We check using \textsc{Pythia} simulations and find that the distribution formed by weighing the pure quark and gluon distributions from $gg \to q\bar{q}$ and $gg \to gg$ processes with the fraction $f_i^{\text{NLL}}$ agrees well with the full simulation. For models 1 and 2 in FIG.~\ref{fig:AA} we choose the following parameters:
\begin{align}
  \text{model 1:} \;\;\;& a = 120 \;\text{GeV},\; b = +1/3 \nn\\
  \text{model 2:} \;\;\;& a = 120 \;\text{GeV},\; b = -1/2.
\end{align}

Also, we demonstrate that different quark and gluon jet fractions can give the same subtracted cumulants, as shown in FIG. \ref{fig:appA}. For simplicity, we assume that the cumulants $\langle \tau\rangle$ of pure quark and pure gluon jets depend linearly on jet $p_T$, and the subtracted cumulants  $\Delta_\tau$ are defined in \eq{tau-gfrac}. The left panel shows the cumulants corresponding to different quark and gluon jet fractions: pure quark, pure gluon and two interpolations between pure quark or gluon across jet $p_T$, as well as cases A and B that sit between the pure quark and gluon cases. The right panel shows the subtracted cumulants. We can clearly see that cases A and B give the same subtracted cumulant, and that the cases of pure quark and gluon jets do not represent extreme values of subtracted cumulants.

\bibliographystyle{apsrev4-1}
\bibliography{bibliography}

\end{document}